\documentclass[aps,nofootinbib,amsmath,prd,twocolumn,showpacs,superscriptaddress,groupedaddress]{revtex4-1}

\usepackage[american]{babel}
\usepackage{amsfonts}
\usepackage{amsmath}
\usepackage{amssymb}
\usepackage{epsfig}
\usepackage{latexsym}
\usepackage{paralist}
\usepackage{fancyhdr}
\usepackage{graphicx}
\usepackage[vcentermath]{youngtab}
\usepackage{young}
\usepackage{ytableau}
\usepackage{etex}
\usepackage{float}
\usepackage{slashed}
\usepackage{xcolor}
\usepackage{soul}
\usepackage{dcolumn} 
\usepackage{bm}
\makeatletter\AtBeginDocument{\let\@elt\relax}\makeatother

\def\bea{\begin{eqnarray}} 
\def\eea{\end{eqnarray}}
\def\be{\begin{equation}} 
\def\ee{\end{equation}} 
\def\ba{\begin{array}}
\def\ea{\end{array}}

\def\be{\begin{equation}}
\def\ee{\end{equation}}
\def\bea{\begin{eqnarray}}
\def\eea{\end{eqnarray}}

\def\Tr{\mbox{Tr}}

\usepackage{amsmath}

\begin{document}

\title{Composite higher derivative operators in $d=2+\epsilon$ dimensions \\ and the spectrum of asymptotically safe gravity}

\author{Riccardo Martini}
\email{riccardo.martini@pi.infn.it}
\affiliation{INFN - Sezione di Pisa, Largo Bruno Pontecorvo 3, 56127 Pisa, Italy}

\author{Dario Sauro}
\email{dario.sauro@phd.unipi.it}
\affiliation{
Universit\`a di Pisa, Largo Bruno Pontecorvo 3, 56127 Pisa, Italy}
\affiliation{INFN - Sezione di Pisa, Largo Bruno Pontecorvo 3, 56127 Pisa, Italy}

\author{Omar Zanusso}
\email{omar.zanusso@unipi.it}
\affiliation{
Universit\`a di Pisa, Largo Bruno Pontecorvo 3, 56127 Pisa, Italy}
\affiliation{INFN - Sezione di Pisa, Largo Bruno Pontecorvo 3, 56127 Pisa, Italy}

\begin{abstract}
%
We discuss the renormalization of Einstein-Hilbert's gravity in $d=2+\epsilon$ dimensions. We show that the application of the path-integral approach leads naturally to scheme- and gauge-independent results on-shell, but also gives a natural notion of quantum metric off-shell, which is the natural argument of the effective action, even at the leading order in perturbation theory. The renormalization group of Newton's constant is consistent with the asymptotic safety scenario for quantum gravity in that it has an ultraviolet relevant fixed point. We extend the approach to the analysis of curvature square operators, understood as composites operators, which allows for the determination of the spectrum of scaling operators at the scale invariant fixed point. The analysis suggests that there is one operator that becomes relevant close to $d=4$ dimensions, while other operators previously found in the literature are either marginal or trivial on-shell.
\end{abstract}

\maketitle

\renewcommand{\thefootnote}{\arabic{footnote}}
\setcounter{footnote}{0}

\section{Renormalized metric's action} \label{sect:introduction}

The asymptotic safety conjecture is based on the premise that Einstein's gravity, if seen as a quantum field theory of the metric tensor, is ultraviolet complete thanks to the presence of a suitable fixed point of the renormalization group \cite{Weinberg:1976xy}. The ultraviolet fixed point must be non-Gaussian in order to circumvent the established fact that the Einstein-Hilbert action is not perturbatively renormalizable \cite{Goroff:1985th}. This was confirmed in a seminal paper using background and Wilsonian renormalization group methods \cite{Reuter:1996cp}, on which most of the recent literature of the topic is now based.

The combined application of background and Wilsonian methods, however, often comes at the price of having to deal with an effective action that is scheme- and gauge-dependent, that is one of the most pressing proplems with which the proposal has to deal \cite{Bonanno:2020bil}. This happens because the effective action is off-shell and the gauge-fixed symmetry is nonlinear. In fact, it is not quite clear how to go on-shell given the approximations that are usually involved in splitting the metric over an arbitrary background and fluctuations on it. Recently, new developments in extracting the on-shell physics were obtained in the context of Wilsonian renormalization and its application to quantum gravity \cite{Baldazzi:2021ydj, Baldazzi:2021orb}.

A way around the problem would be to address scheme- and gauge-dependence in a setting in which perturbation theory is applicable, e.g., one in which
the ultraviolet fixed point is still perturbative. This setting is provided by gravity in $d=2+\epsilon$ dimensions, that exhibits an order-$\epsilon$ ultraviolet
fixed point for Newton's constant motivating the asymptotic safety since its inception \cite{Weinberg:1980gg}. In fact, this setting was the first testing ground of asymptotic safety \cite{Kawai:1989yh}. The obvious limitation is that the continuation to $d=4$ requires $\epsilon=2$, which is certainly outside the validity of perturbation theory, which is why it should be regarded as a complementary approach to the functional one \cite{Martini:2021lcx}.
{Furthermore, pure truly-two-dimensional gravity is topological in nature, because the curvature scalar is also the Euler's density, therefore it is unclear if the continuation of such model is actually useful in $d=4$ where there actually are propagating degrees of freedom among other things.
}

In a relatively recent paper, Ref.~\cite{Martini:2021slj}, the approach in $d=2+\epsilon$
was reconsidered with an ample discussion on the diffeomorphisms symmetry, its gauge fixing, and the parametric dependence induced by the background splitting.
{The analysis of Ref.~\cite{Martini:2021slj} also shows how to circumvent the topological nature in $d=2$, and we clarify its logic in more detail below.
}
The starting point is a naive dimensional continuation of the Einstein-Hilbert action
\begin{equation}\label{eq:einstein-hilbert}
\begin{split}
 S_E[g] &= \int{\rm d}^dx \sqrt{g}\Bigl\{ g_0-g_1 R\Bigr\}\,,
\end{split}
\end{equation}
where $g_1$ is the inverse of Newton's constant, modulo a normalization, and $g_0$
is a cosmological constant-like contribution. In two dimensions, $g_1$ is dimensionless and its inverse is a suitable coupling to construct the perturbative expansion, while $g_0$ gives a dimension two deformation that is going to be useful when going on-shell.
In general dimension $d$, we have that $\Lambda = g_0/g_1$ is a dimension two constant (the cosmological constant) and this is going to play an important role in the next section
because it could mix with dimensionful composite operators.
{The reader familiar with Ref.~\cite{Martini:2021slj} can skip to the discussion preceding Eq.~\eqref{eq:sct-comp}, but we recommend to read the following more in-depth discussion.
}

The scalar curvature term of \eqref{eq:einstein-hilbert} is topological in two dimensions, so there are two distinct ways to implement the diffeomorphisms symmetry at the quantum level given the discontinuity of the number of degrees of freedom of $g_{\mu\nu}$. The discussion of Ref.~\cite{Martini:2021slj} strongly suggests that the simplest realization of the (infinitesimal) diffeomorphisms
\begin{equation}\label{eq:diff-action}
\begin{split}
 \delta_\xi g_{\mu\nu} &= {\cal L}_\xi g_{\mu\nu} = \nabla_\mu \xi_\nu+ \nabla_\nu \xi_\mu\,,
\end{split}
\end{equation}
is the one best suited for analytic continuation to general $d$.\footnote{
Earlier results
were based on an isomorphic realization of the diffeomorphisms group
that isolates the conformal/dilaton mode
like in string theory \cite{Kawai:1989yh}.
In Ref.~\cite[Sect.~V]{Martini:2021slj} it was shown that the dilaton realization is not suitable for continuation in $d$, because results are gauge-dependent even on-shell so dilaton-based proposals, e.g., \cite{Elizalde:1995gw}, are not pertinent to this paper. A qualitative reason for the failure of the dilaton realization is because it requires the cancellation of the central-charge of the trace anomaly for consistency, but such anomaly is well-defined only in $d=2$.
}
The algebra of the group is isomorphic to the one of Lie-brackets, $\left[\delta_{\xi_1},\delta_{\xi_2}\right] = \delta_{\left[\xi_1,\xi_2\right]}$.

A path-integral for the action \eqref{eq:einstein-hilbert} should be constructed integrating over all possible metrics.
In order to manifestly maintain covariance, we use the background field method introducing an arbitrary metric $\overline{g}_{\mu\nu}$ and fluctuations $h_{\mu\nu}$
\begin{equation}\label{eq:linear-split}
\begin{split}
 g_{\mu\nu} &= \overline{g}_{\mu\nu} + h_{\mu\nu}
+\frac{\lambda}{2} h_{\mu\rho} \overline{g}^{\rho\theta} h_{\theta\nu} +{\cal O}(h^3)
\,,
\end{split}
\end{equation}
and integrate the path-integral over the fluctuations. Notice that the above split 
introduces the constant parameter $\lambda$, which is \emph{fundamental} for us to test the scheme dependence of the procedure in the rest of the paper. Further orders in the expansion could be parametrized too, but they will not play a role in the following.

The manifest symmetry is the one for which the background version of \eqref{eq:diff-action} is realized. The symmetry that is gauge-fixed, instead, is the one for which the background is kept fixed, implying that $h_{\mu\nu}$ transforms nonlinearly
\begin{equation}\label{eq:hmunu-transf-linear}
\begin{split}
 \delta_\xi h_{\mu\nu} =
  \overline{g}_{\rho\nu}\overline{\nabla}_{\mu} \xi^{\rho}+\overline{g}_{\rho\mu}\overline{\nabla}_{\nu} \xi^{\rho} + {\cal O}(h)\,.
\end{split}
\end{equation}
The rhs of Eq.~\eqref{eq:hmunu-transf-linear} is nonlinear in $h_{\mu\nu}$ and all orders can be computed with some work, but they are not needed in what follows.
A simple gauge-fixing is the Feynman-deDonder one, which we choose
\begin{equation}\label{eq:dedonder}
\begin{split}
 S_{\rm gf} &= \frac{1}{2} \int {\rm d}^dx \sqrt{\overline{g}}\,\overline{g}^{\mu\nu} F_\mu F_\nu\,,
 \\
 F_\mu & = \overline{\nabla}^\rho h_{\rho\mu} - \frac{\beta}{2} \overline{g}^{\rho\theta}\overline{\nabla}_\mu h_{\rho\theta}\,,
\end{split}
\end{equation}
that it introduces an additional parameter $\beta$. In Ref.~\cite{Martini:2021slj}
the parameter is chosen as $\beta=1+\delta\beta$ with $\delta\beta\ll 1$ for simplicity. The independence on $\delta\beta$ is not enough to fully establish gauge-independence of the final results (it is necessary but not sufficient condition), which is why the $\lambda$ dependence of \eqref{eq:linear-split} is most important in the following.
From \eqref{eq:dedonder} and \eqref{eq:hmunu-transf-linear}, it is easy to determine the ghosts action
\begin{equation}\label{eq:ghost}
\begin{split}
 S_{\rm gh} &= \int {\rm d}^dx \sqrt{\overline{g}} \, \overline{c}^\mu \left.\delta_\xi F_\mu\right|_{\xi\to c} \,.
\end{split}
\end{equation}

\subsection{Path-integral}

We have all the ingredients to construct the Euclidean path-integral over the background
\begin{equation}\label{eq:pi}
\begin{split}
 Z[j] &= \int {\rm D}h \, {\rm e}^{-S_E[\overline{g}+h]
 -S_{\rm gf}-S_{\rm gh} + j\cdot h}\,,
\end{split}
\end{equation}
where the measure also contains the integration over the ghosts,
and the effective action
\begin{equation}\label{eq:ea}
\begin{split}
 {\rm e}^{-\Gamma} &= \int {\rm D}h \, {\rm e}^{-S_E[\overline{g}+h]
 -S_{\rm gf}-S_{\rm gh} + \frac{\delta\Gamma}{\delta \langle h\rangle}\cdot(h-\langle h\rangle)}\,.
\end{split}
\end{equation}
Using standard manipulations, it is straightforward to obtain the one-loop effective action
\begin{equation}\label{eq:gamma}
\begin{split}
 \Gamma &= S_E + \frac{1}{2} \Tr \log {\cal O}^{\beta,\lambda}_h - \Tr \log {\cal O}^{\beta,\lambda}_{\rm gh}\,,
\end{split}
\end{equation}
where ${\cal O}^{\beta,\lambda}_h$ and ${\cal O}^{\beta,\lambda}_{\rm gh}$ are the Hessians of $h_{\mu\nu}$ and ghosts, respectively. At this stage, everything depends parametrically on the gauge and the split through $\beta$ and $\lambda$.

The effective action $\Gamma$ is naively a functional of
the expectation value $\langle h_{\mu\nu} \rangle$ and also depends on $\overline{g}_{\mu\nu}$, i.e., $\Gamma=\Gamma[\overline{g}_{\mu\nu},\langle h_{\mu\nu} \rangle]$.
The double dependence is expected to merge in a suitable way according to the Ward identity of a split symmetry induced by all combined transformations preserving \eqref{eq:linear-split}. This symmetry is not anomalous if we use a regularization prescription that is cutoff independent, which is why we choose dimensional regularization, and this represents a main departure point from the Wilsonian approaches as in \cite{Reuter:1996cp}. The red herring, to be kept in mind for the following, is that we are tempted to assume that $\Gamma$ is a natural function of a single ``quantum'' metric, $\Gamma=\Gamma[\langle g_{\mu\nu}\rangle]$, where $\langle g_{\mu\nu}\rangle= \overline{g}_{\mu\nu}+\langle h_{\mu\nu}\rangle +\dots$ and the dots hide corrections that should be computed through external legs of $\langle h_{\mu\nu}\rangle$.

The evaluation of the divergent part of $\Gamma$ can be carried out in the limit $\langle h_{\mu\nu} \rangle=0$, which is the most important feature of the background method. In this limit ${\cal O}^{\beta,\lambda}_h$ and ${\cal O}^{\beta,\lambda}_{\rm gh}$
are differential operators that depend on covariant derivatives and curvatures of the background. Furthermore, the differential operators are of Laplace-type, except for the insertion caused by $\delta\beta$ which can be expanded because of our choice.

We illustrate this considering ${\cal O}^{\beta,\lambda}_h$ first, and a similar discussion applies to ${\cal O}^{\beta,\lambda}_{\rm gh}$.
It is expanded
\begin{equation}\label{eq:hessian-h-full}
\begin{split}
 {\cal O}^{\beta,\lambda}_h = {\cal O}^{\lambda}_h + \delta\beta \, \hat{P}^\lambda(\overline{\nabla})\,,
\end{split}
\end{equation}
where
\begin{equation}
\begin{split}
 {\cal O}^{\lambda}_h = -\overline{\nabla}^2 K + U^\lambda\,,
\end{split}
\end{equation}
$K$ and $U^\lambda$ are metric-dependent endomorphisms acting on symmetric tensor fields, and $\hat{P}^\lambda$ is a differential operator with uncontracted derivatives (all indices have been suppressed for brevity). It is useful to display $K$ explicitly
\begin{equation}\label{eq:K}
\begin{split}
K^{\mu\nu}{}_{\rho\theta} = \delta^{(\mu}_\rho\delta^{\nu)}_\theta - \frac{1}{2}\overline{g}^{\mu\nu}\overline{g}_{\rho\theta}\,,
\end{split}
\end{equation}
so that it is clear that the Green function of ${\cal O}^{\lambda}_h$,
defined as ${\cal O}^{\lambda}_{h,x}\cdot {\cal G}_{h,xy}=\delta_{xy}$,
is proportional to the inverse of \eqref{eq:K}
\begin{equation}\label{eq:Kinv}
\begin{split}
K^{-1}{}^{\mu\nu}{}_{\rho\theta} = \delta^{(\mu}_\rho\delta^{\nu)}_\theta - \frac{1}{d-2}\overline{g}^{\mu\nu}\overline{g}_{\rho\theta}\,,
\end{split}
\end{equation}
and contains a kinematical pole $1/(d-2)$ caused by the fact that the Einstein-Hilbert action is topological in two dimensions and the number of degrees of freedom changes discontinuously across $d=2$. We stress that such a pole appears in \eqref{eq:Kinv} regardless of the choice of gauge fixing.
While it looks exactly like a pole of dimensional regularization, the kinematical pole presents a danger if subtracted in renormalization, as demonstrated
in the inconsistencies found in Ref.~\cite{Jack:1990ey}. The reason is that in dimensional regularization poles should come in the form $\mu^{\epsilon}/\epsilon$ for $\epsilon=(d-2)$, or else higher order poles will not cancel correctly in subdivergences \cite{Osborn:1987au}. We elaborate more on this in the next sections and in Appendix~\ref{sect:appendix}.

\subsection{Covariant regularization of the graviton loops and $d$-dependence}
\label{sect:graviton-loops}

The presence of {the kinematical pole is crucial for understanding some choices in the renormalization scheme chosen in the paper, so it is worth discussing it more now.
One may expect that the pole $1/(d-2)$ should be canceled with counterterms, just like the poles of dimensional regularization, in fact
in the past the pole has sparked a debate that has lead to different conclusions \cite{Jack:1990ey,Aida:1996zn}. However, this pole should not be subtracted with counterterms, or else there are inconsistencies in the renormalization prescription
which become dangerous at two loops \cite{Jack:1990ey}, where there is a delicate balance between higher order poles and subdivergences. By extension, this problem manifests in the renormalization of composite operators, which is the subject of Sect.~\ref{sect:hd-operators} of this paper. The solution is actually simple, because it was shown in Ref.~\cite{Martini:2021slj} that there is no need to subtract the pole, since it \emph{cancels} in physical on-shell results, which we confirm below for the renormalisation of both $\Gamma$ and composite operators. 
}

The computation is most easily performed using standard heat kernel techniques with the Seeley-deWitt expansion, and we continue the illustration using ${\cal O}^{\beta,\lambda}_h$.
The second term of \eqref{eq:gamma} is expressed as
\begin{eqnarray}\label{eq:gamma-h-contribution}
\frac{1}{2} \Tr \log {\cal O}^{\beta,\lambda}_h
&=& -\frac{1}{2} \int \frac{{\rm d}s}{s} \, \Tr\,  {\rm e}^{- s {\cal O}^{\beta,\lambda}_h}
\\
&=& -\frac{1}{2} \int \frac{{\rm d}s}{s} \, \Tr\Bigl(
{\rm e}^{- s {\cal O}^{\lambda}_h}
- s \, \delta\beta \, \hat{P}^\lambda {\rm e}^{- s {\cal O}^{\lambda}_h}
\Bigr)\,,\nonumber
%
\end{eqnarray}
and ${\cal O}^{\lambda}_h= {\cal O}^{\beta,\lambda}_h|_{\beta=1}$
Both the trace of the heat kernel, i.e., $\Tr\exp(- s {\cal O}^{\lambda}_h)$,
and the heat kernel with derivative insertions, i.e., $\Tr\hat{P}^\lambda\exp(- s {\cal O}^{\lambda}_h)$,
admit asymptotic expansions in terms of well-known computable coefficients \cite{Groh:2011dw}.\footnote{%
A technical point: in Ref.~\cite{Groh:2011dw} the coefficients of the expansion are normalized without the square root of the Van Vleck determinant, as was suggested in Ref.~\cite{Anselmi:2007eq}. For the computations of this paper we have used the more conventional normalization of Refs.~\cite{Jack:1983sk,Martini:2018ska}. The final results are independent of this choice of course.
}
Schematically
\begin{equation}\label{eq:sdw-expansions-schematically}
\begin{split}
&\Tr\, {\rm e}^{- s {\cal O}^{\lambda}_h} \sim \frac{1}{s^{d/2}}\Bigl\{
a_0(\overline{g};\lambda,d) +s \, a_1(\overline{g};\lambda,d)+\cdots
\Bigr\}\,, 
\\
&\Tr\, \hat{P}^\lambda {\rm e}^{- s {\cal O}^{\lambda}_h} \sim
\frac{1}{s^{d/2+1}}\Bigl\{
\hat{a}_0(\overline{g};\lambda,d) +\cdots
\Bigr\}\,, 
\end{split}
\end{equation}
{where the coefficients $a_n$ and $\hat{a}_n$ of the expansions in the numerators
of the rhs are constructed with the metric $\overline{g}_{\mu\nu}$
and its curvatures. Dimensional analysis tells us that $a_n \sim {\cal R}^n$ and $\hat{a}_n\sim {\cal R}^n$, where the symbol ${\cal R}={\cal R}[g]$ weighs as a Riemannian curvature or two covariant derivatives \cite{Groh:2011dw}. The coefficients are given in Appendix~\ref{sect:appendixHK}.

It is crucial to realize at this point
that the dimensionality $d$ appears
in two clearly identifiable and distinct ways in the asymptotic expansion:
$d$ governs the leading powers of the expansions of Eqs.~\eqref{eq:sdw-expansions-schematically}, but it also appears \emph{parametrically} within the coefficients
$a_n$ and $\hat{a}_n$
of the Seeley-deWitt expansion through the aforementioned kinematical pole
and the metric itself (e.g., $\overline{g}^\mu{}_\mu=d$). The latter $d$-dependence is the one properly
related to the number of (off-shell) propagating degrees of freedom and, in fact, it is the one accounting for the discontinuity in $d=2$ caused by the kinematical pole itself.

Since our aim is to continue the results to $d>2$ and eventually to $d=4$, it would be desirable to propagate the correct off-shell degrees of freedom of $h_{\mu\nu}$ in general $d$, but \emph{at the same time} use a continuation of $d$ which allows to
isolate a finite number of counterterms and make the theory finite by subtraction.
In the covariant generalization of dimensional regularization one finds a finite number of counterterms by continuing the $d$ that appears in the \emph{leading} power of $s$
in \eqref{eq:sdw-expansions-schematically}. For example, close to $d=2$, the theory is
renormalizable and has two counterterms that need subtraction, which appear as
those scaling as $s^{-d/2-1}$ when inserting \eqref{eq:sdw-expansions-schematically} in \eqref{eq:gamma-h-contribution}. Those are the universal poles, i.e., the ultraviolet logarithmic divergences \cite{Jack:1983sk}.

The point made in Ref.~\cite{Martini:2021slj} is that the covariant diagrammatics is always convergent if the leading $d$ dependence is continued away from $d=2$ in the complex plane to a value with $\Re(d)<2$, where all modes, including the conformal one, become stable. At the same time, we are free to leave intact the parametric $d$-dependence in the coefficients of the Seeley-deWitt expansion, ensuring that the correct number of off-shell degrees of freedom are propagating in any $d$.
For this reason,
we introduce the analytic continuation $d=2+\epsilon$ \emph{only} on the leading powers. While this choice may seem arbitrary at this stage, it is nontrivially confirmed to be consistent through the several cancellations that we show below.
In short, the regulated heat kernels are
\begin{equation}\label{eq:sdw-expansions-schematically-regulated}
\begin{split}
&\Tr\, {\rm e}^{- s {\cal O}^{\lambda}_h}|_{\rm reg} \sim -\frac{1}{s^{1+\epsilon/2}}\Bigl\{
a_0(\overline{g};\lambda,d) +s \,a_1(\overline{g};\lambda,d)+\cdots
\Bigr\}\,, 
\\
&\Tr\, \hat{P}^\lambda {\rm e}^{- s {\cal O}^{\lambda}_h}|_{\rm reg} \sim
-\frac{1}{s^{2+\epsilon/2}}\Bigl\{
\hat{a}_0(\overline{g};\lambda,d) +\cdots
\Bigr\}\,, 
\end{split}
\end{equation}
(compare with Eq.~\eqref{eq:sdw-expansions-schematically}) and they can be used in \eqref{eq:gamma-h-contribution} to find the contributions of $h_{\mu\nu}$ to the regulated effective action at one loop
\begin{eqnarray}\label{eq:gamma-h-contribution-regulated}
\frac{1}{2} \Tr \log {\cal O}^{\beta,\lambda}_h |_{\rm reg}
&=& \frac{\mu^{\epsilon}}{\epsilon}\Bigl\{
a_1(\overline{g};\lambda,d)
-\delta\beta \, \hat{a}_1(\overline{g};\lambda,d)
\Bigr\}\nonumber
\\
&& + ~ {\rm finite}\,. 
%
\end{eqnarray}
Notice that, as a result of our procedure, the divergences have retained the $d$-dependence caused by the number of propagating degrees of freedom, even though the pole in $\epsilon$ is close to $d=2$.

\subsection{Covariant regularization of the ghost loops}

We can now follow the same strategy to regulate the ghost contribution of \eqref{eq:gamma}. We start by splitting the differential operator ${\cal O}^{\beta,\lambda}_{\rm gh}$ as
\begin{align}
{\cal O}^{\beta,\lambda}_{\rm gh} = {\cal O}^{\lambda}_{\rm gh} + \delta\beta\;\hat{Q}^\lambda(\bar{\nabla})\,.
\end{align}
Notice that both the operators ${\cal O}^{\lambda}_{\rm gh}$ and $\hat{Q}^\lambda$ in general contain interactions between the graviton and the ghosts, however these do not contribute to the one-loop background effective action because of the limit $\langle h_{\mu\nu}\rangle=0$. Furthermore, ${\cal O}^{\lambda}_{\rm gh} =  {\cal O}^{\beta\lambda}_{\rm gh}|_{\beta=1} \sim -\overline{\nabla}^2+{\rm Ric}$. The trace of the logarithm of ${\cal O}^{\beta,\lambda}_{\rm gh}$ can then be expressed using a proper time representation as
\begin{align}
\label{eq:ghostTrLog}
\Tr\log {\cal O}^{\beta,\lambda}_{\rm gh} =& - \int \frac{{\rm d}s}{s} \, \Tr\,  {\rm e}^{- s {\cal O}^{\beta,\lambda}_{\rm gh}}
\\
=& - \int \frac{{\rm d}s}{s} \, \Tr\Bigl(
{\rm e}^{- s {\cal O}^{\lambda}_{\rm gh}}
- s \, \delta\beta \, \hat{Q}^\lambda {\rm e}^{- s {\cal O}^{\lambda}_{\rm gh}}
\Bigr)\,.\nonumber
\end{align}
Similarly to Sect.~\ref{sect:graviton-loops}, we provide the asymptotic expansion of the two traces of \eqref{eq:ghostTrLog} as
\begin{equation}
\begin{split}
&\Tr\,{\rm e}^{- s {\cal O}^{\lambda}_{\rm gh}} \sim \frac{1}{s^{d/2}} \Big\{b_0(\bar{g};\lambda, d)+s\,b_1(\bar{g};\lambda, d)+\dots\Big\}\,,\\
&\Tr\,\hat{Q}^\lambda\,{\rm e}^{- s {\cal O}^{\lambda}_{\rm gh}} \sim \frac{1}{s^{d/2+1}}\Big\{\hat{b}_0(\bar{g};\lambda, d)+\dots\Big\}\,,
\end{split}
\end{equation}
with new heat-kernel coefficients $b_n$ and $\hat{b}_n$.
The regularization follows the scheme described in Sect.~\ref{sect:graviton-loops}
of continuing only the leading power of $s$, as in Eq.~\eqref{eq:sdw-expansions-schematically-regulated}. We obtain
\begin{eqnarray}\label{eq:gamma-ghost-contribution-regulated}
\Tr \log {\cal O}^{\beta,\lambda}_{\rm gh} |_{\rm reg}
&=& 2\frac{\mu^{\epsilon}}{\epsilon}\Bigl\{
b_1(\overline{g};\lambda,d)
-\delta\beta \, \hat{b}_1(\overline{g};\lambda,d)
\Bigr\}\nonumber
\\
&& + ~ {\rm finite}\,. 
%
\end{eqnarray}
The coefficients are given in Appendix~\ref{sect:appendixHK}.

\subsection{Renormalization of $\Gamma$}

Now we combine the results \eqref{eq:gamma-h-contribution-regulated} and \eqref{eq:gamma-ghost-contribution-regulated} in the expression of $\Gamma$ given in \eqref{eq:gamma} to obtain the regulated effective action as
\begin{equation}\label{eq:gamma-regulated}
\begin{split}
 \Gamma|_{\rm reg} =& S_E + \frac{1}{2} \Tr \log {\cal O}^{\beta,\lambda}_h|_{\rm reg} - \Tr \log {\cal O}^{\beta,\lambda}_{\rm gh}|_{\rm reg}\\
 =& S_E + \frac{\mu^{\epsilon}}{\epsilon}\left\{a_1-2b_1+\delta\beta\left(2\hat{b}_1-\hat{a}_1\right)\right\}\\
 &+ {\rm finite}\,.
\end{split}
\end{equation}
The divergent part $\Gamma_{\rm div}$, i.e., the $1/\epsilon$
poles of $\Gamma|_{\rm reg}$, should be subtracted with the appropriate redefinitions of the bare couplings in $S_E$. They inherit the $d$- and $\lambda$-dependences from the heat kernel coefficients.
A qualitative discussion on the interplay between gravity and dimensional regularization is given in Appendix~\ref{sect:appendix}, while the explicit forms of the coeffiecients $a_1$, $\hat{a}_1$, $b_1$ and $\hat{b}_1$ are all given in Appendix~\ref{sect:appendixHK}.
}

The result of the above procedure
gives the divergent part of the effective action, which depends only on the background because, we stress once more, it was computed in the limit $\langle h_{\mu\nu} \rangle=0$ \cite{Martini:2021slj}.
The most convenient way to write the result is to use the background scalar curvature and the equations of motion of the bare action,
\begin{equation}\label{eq:sct-comp}
\begin{split}
 & \Gamma_{\rm div}
  = -\frac{\mu^{\epsilon}}{\epsilon} \int {\rm d}^d x \sqrt{\overline{g}} \Bigl\{
  A_d \overline{R} + J^{\beta,\lambda}_{\mu\nu}  E[\overline{g}]^{\mu\nu} \Bigr\}
\,,\\
 & E[\overline{g}]^{\mu\nu} = \overline{G}^{\mu\nu} + \frac{g_0}{2g_1}\overline{g}^{\mu\nu} = \overline{G}^{\mu\nu} + \frac{\Lambda}{2}\overline{g}^{\mu\nu}\,,
\end{split}
\end{equation}
where $E[\overline{g}]^{\mu\nu}$ are the equations of motion evaluated on the background metric,
$\overline{G}_{\mu\nu}=\overline{R}_{\mu\nu}-\frac{1}{2}\overline{R}\, \overline{g}_{\mu\nu}$ is the background's Einstein tensor,
the coefficient $A_d$ is just a scalar that depends parametrically \emph{only} on $d$, and instead
$J^{\beta,\lambda}_{\mu\nu}$ depends on gauge- and scheme-parameters
\begin{eqnarray}\label{eq:sct-coeff}
 A_d&=&\frac{1}{4\pi}\frac{36+3d-d^2}{12 }\,,
\\
 J^{\beta,\lambda}_{\mu\nu} &=& \frac{\overline{g}_{\mu\nu}}{4\pi} \Bigl\{
 \frac{d^2-d-4 }{2 (d-2)}\lambda-\delta\beta  \left(2+\frac{2 \lambda
   }{d-2}\right)-d-1
\Bigr\} \,.\nonumber
\end{eqnarray}
In essence, the nonzero value for $g_0$ in \eqref{eq:einstein-hilbert} was introduced precisely because it allows to have $\overline{R}\neq 0$ and unambiguously extract the coefficient $A_d$ on-shell. Most notably and quite nontrivially, $J^{\beta,\lambda}_{\mu\nu}$ contains all the ``unwanted'' dependencies, including the kinematical pole, suggesting a meaningful $d\to 2$ limit despite the discontinuity of the number of degrees of freedom.

{The immediate conclusion of \eqref{eq:sct-comp} is that all the unwanted parameters and the kinematical pole in $J^{\beta,\lambda}_{\mu\nu}$ decouple on-shell. The remaining coefficient $A_d$ is naturally subtracted by renormalizing the coupling $g_1$, so it is used to obtain the beta function of the dimensionless Newton's constant $G=\mu^{\epsilon}/g_1$
\begin{equation}\label{eq:betaG-JJ-epsilon}
\begin{split}
 \beta_G&= \epsilon G - A_d G^2\,,
\end{split}
\end{equation}
which is a universal, gauge- and scheme-independent result, thanks to the fact that all the unwanted dependencies decoupled on-shell through $J^{\beta,\lambda}_{\mu\nu}$ in \eqref{eq:sct-comp}. Notice that we have now moved to the renormalized coupling $G$, expressed in units of the renormalization group scale $\mu$.

Even though we have kept the $d$-dependence of the counterterm through $A_d$ and that of dimensional regularization through $\epsilon=(d-2)$ separate, it is important to realize that, having regulated the theory close to two dimensions, the perturbative series is truly consistent
when they are identified. In this case, the beta function \eqref{eq:betaG-JJ-epsilon} becomes
\begin{equation}\label{eq:betaG-JJ}
\begin{split}
 \beta_G&= (d-2) G - \frac{36+3d-d^2}{48 \pi } G^2\,,
\end{split}
\end{equation}
which confirms a result of Ref.~\cite{Falls:2017cze} obtained with a cutoff prescription, and results of Ref.~\cite{Bastianelli:2022pqq} in which the $d$-dependence was kept through the computations.
The solution of $\beta_G=0$ gives a $G^*\sim O(d-2)$ ultraviolet fixed point.
Extrapolation to finite values of $\epsilon=(d-2)$ gives $G^*$ as a function of $d$
\begin{equation}\label{eq:uv-fp}
\begin{split}
G^*=\frac{48\pi (d-2)}{36+3d-d^2}\,,
\end{split}
\end{equation}
which exists up to $d\approx 7.7$, so ``safely'' above $4$, although further corrections should be computed
since this is an extrapolation to high values of $\epsilon$.

\subsection{Renormalization of $\langle h_{\mu\nu} \rangle$ and quantum metric}

One observation, that is going to be very important in the next section, comes from realizing that \eqref{eq:sct-comp} is also telling us what is the wavefunction renormalization of the full quantum metric.
Naively, the quantum metric is the natural \emph{single} argument of $\Gamma_{\rm ren}$ and is expected to be of the form $\langle g_{\mu\nu} \rangle \simeq \overline{g}_{\mu\nu} +\langle h_{\mu\nu} \rangle +\cdots$, which generalizes the split of Eq.~\eqref{eq:linear-split}. However, in the background field method, $\langle h_{\mu\nu} \rangle$ is subject to a nonlinear renormalization because it transforms nonlinearly. The nonlinear renormalization is generally computed from covariant diagrams with one external fluctuation's leg \cite{Howe:1986vm,Osborn:1987au}.
In other words, $\langle h_{\mu\nu} \rangle$ is not the ``true'' renormalized fluctuation,
which is the natural expansion of the argument of the path-integral of the effective action over $\overline{g}_{\mu\nu}$.

Even though we have been working in the limit $\langle h_{\mu\nu} \rangle=0$, there is a simpler way to reconstruct its \emph{leading} renormalization. Subtracting the counterterms, we have that \eqref{eq:sct-comp} can be rewritten in terms of the renormalized coupling by moving the $J^{\beta,\lambda}_{\mu\nu}$ term to the rhs
\begin{equation}\label{eq:gamma-almost-renormalized}
\begin{split}
 \Gamma_{\rm ren}[\overline{g}]+\frac{\mu^{\epsilon}}{\epsilon}J^{\beta,\lambda}_{\mu\nu}\frac{\delta\Gamma_{\rm ren}[\overline{g}]}{\delta \overline{g}_{\mu\nu}}  &
  =  -\frac{1}{G} \int {\rm d}^d x \sqrt{\overline{g}}
  \overline{R} 
\,,
\end{split}
\end{equation}
to the leading order in $\hbar$ (which is always implicit in our notation), having identified the equations of motion of $\Gamma_{\rm ren}$ and the renormalized action in the limit $\langle h_{\mu\nu}\rangle=0$.
A trivial manipulation is to rewrite the lhs of Eq.~\eqref{eq:gamma-almost-renormalized} as
\begin{equation}\label{eq:gamma-almost-almost-renormalized}
\begin{split}
 \left.
 \Gamma_{\rm ren}[\overline{g}]+\frac{\mu^{\epsilon}}{\epsilon}J^{\beta,\lambda}_{\mu\nu}\frac{\delta\Gamma_{\rm ren}[\overline{g}+\langle h\rangle ]}{\delta \langle h_{\mu\nu}\rangle}
 \right|_{\langle h \rangle=0}
\,,
\end{split}
\end{equation}
which is the Taylor expansion to the leading order in $\hbar$ on the off-shell renormalized action with a new argument
\begin{equation}\label{eq:gamma-almost-almost-almost-renormalized}
\begin{split}
 \Gamma_{\rm ren}\left[\overline{g}+\langle h \rangle + \frac{\mu^{\epsilon}}{g_1\epsilon} J^{\beta,\lambda}_{\mu\nu} \right]
\,.
\end{split}
\end{equation}
This shows that the renormalized action $\Gamma_{\rm ren}$ is actually a natural function of a renormalized metric that is \emph{not} $\langle g_{\mu\nu} \rangle =\overline{g}_{\mu\nu}+\langle h_{\mu\nu}\rangle +\cdots $, even in the limit $\langle h_{\mu\nu} \rangle=0$. To the leading order in perturbation theory
the renormalized metric is
\begin{equation}\label{eq:ren-metric-almost}
\begin{split}
 \langle g_{\mu\nu} \rangle  = \overline{g}_{\mu\nu} +\langle h_{\mu\nu}\rangle +\frac{\mu^{\epsilon}}{g_1\epsilon} J^{\beta,\lambda}_{\mu\nu}
 + {\cal O}(\langle h \rangle^2)
\,,
\end{split}
\end{equation}
and we stress that the renormalization is caused by the fluctuation $\langle h_{\mu\nu}\rangle$ and \emph{not} by a renormalization of the background $\overline{g}_{\mu\nu}$.
Since the metric is not a true observable, being gauge-dependent, it is not unexpected to see that its renormalization depends on gauge- and scheme-parameters.

One last manipulation is useful to conclude this section. Since above we have computed everything in the limit $\langle h_{\mu\nu}\rangle=0$, it is convenient to introduce a new background metric $\tilde{g}_{\mu\nu}$ as the same limit of the full quantum metric
\begin{equation}\label{eq:ren-metric}
\begin{split}
 \tilde{g}_{\mu\nu} = \langle g_{\mu\nu} \rangle|_{\langle h \rangle=0} = \overline{g}_{\mu\nu} +\frac{\mu^{\epsilon}}{g_1\epsilon} J^{\beta,\lambda}_{\mu\nu}
\,,
\end{split}
\end{equation}
where the expression for $J^{\beta,\lambda}_{\mu\nu}$ is always given in \eqref{eq:sct-coeff} and the result still depends on the unphysical parameters, because the metric itself is not straightforwardly an observable. This should \emph{not} be regarded as a renormalized background metric, as it would be unphysical, but as the special limit of the renormalized metric defined in Eq.~\eqref{eq:ren-metric-almost}.
Using the (special limit of the) renormalized metric
\begin{equation}\label{eq:gamma-renormalized}
\begin{split}
 \Gamma_{\rm ren}[\tilde{g}] &
  =  -\frac{1}{G} \int {\rm d}^d x \sqrt{\tilde{g}}
  \tilde{R} 
\,,
\end{split}
\end{equation}
which, again, is valid to the leading order in perturbation theory.
This is the first important result of this paper: a finite renormalized action that depends only
on the finite renormalized coupling, denoted $G$, and on the renormalized metric
through
$\tilde{g}_{\mu\nu}$. The use of \eqref{eq:ren-metric} and the second term of the lhs of \eqref{eq:gamma-almost-renormalized} is fundamental to have
a completely finite and manifestly gauge- and scheme-independent renormalized result, in a manner similar to other nonlinear theories \cite{Osborn:1987au},
which corrects the naive assumption discussed after Eq.~\eqref{eq:gamma}.
This is particularly important in the next application
that involves the deformation of the original bare action
through the inclusion of composite operators that require further renormalization,
since the new counterterms must be expressed in terms of $\tilde{g}_{\mu\nu}$ for consistency.

}

\section{Higher derivative operators}\label{sect:hd-operators}

A suitable ultraviolet fixed point for asymptotic safety must satisfy a
few important properties. The first and possibly most important of these is that the dimensionality
of the (ultraviolet) critical surface of the fixed point in the space of couplings
is finite, since this dimensionality is linked to the number of uv relevant operators and hence of undetermined parameter of the theory \cite{Weinberg:1976xy}.
The critical surface can be obtained by studying the linearized renormalization group flow close to the uv fixed point.

In Wilsonian approaches the number of uv relevant operators is generally obtained by truncating the effective action with a controlled number of operators, which may or may not depend on the background. So the flow is projected onto the truncation and the number of relevant operators can be found. This operation, however, shares some limitation with the rest of the framework, even neglecting the cases in which the background specializes the projection. In fact, the renormalization group flow that is generally projected is the one of the off-shell action, which, as we discussed before, is gauge- and parameter-dependent. Furthermore, the spectrum of operators may not be entirely physical given that some scaling operators can be constructed from traces and contractions of the equations of motion.\footnote{%
Having an off-shell functional effective action is not intrinsically bad,
but drawing physical conclusions on the number of (ir)relevant operators is potentially dangerous.
} For an overview of the essential scheme in perturbation theory and its application to composite operators see Ref.~\cite{Anselmi:2012aq}.

For example, the current understanding of the spectrum is that a linear combination of square curvature operators is a relevant operator in $d=4$, while two more combinations are irrelevant \cite{Codello:2007bd}. A similar story might happen for cubic curvature operators \cite{Kluth:2020bdv,Kluth:2022vnq}. In this section we want to show, within the limitations of perturbation theory, that there is one and only one additional operator that is uv relevant and quadratic in the curvatures.

To begin with, we consider a basis for the bare higher derivative operators,
which we could parametrize as
\begin{equation}\label{eq:hd-operators-no-eoms}
\begin{split}
S_{\rm hd}[g] &=
-\int {\rm d}^d x \,\sqrt{g} \Bigl\{ c_1 R_{\mu\nu\rho\sigma}^2  + c_2 R^2
+c_3 R_{\mu\nu}^2\\
& \qquad\qquad\qquad\qquad + c_4 \Lambda R + c_5 \Lambda^2\Bigr\}
\end{split}
\end{equation}
where $R_{\mu\nu}^2=R_{\mu\nu}R^{\mu\nu}$ and $R_{\mu\nu\rho\sigma}^2=R_{\mu\nu\rho\sigma}R^{\mu\nu\rho\sigma}$ and the last two terms are added as potential mixing on the basis of dimensional analysis (recall that $\Lambda=g_0/g_1$ is dimension two).
However, the above parametrization in terms of the couplings $c_i$
is not convenient when going on-shell,
and we do need to go on-shell later on. For this reason, we introduce the linearly related couplings $\alpha_i$ through
\begin{equation}\label{eq:hd-operators}
 \begin{split}
S_{\rm hd}[g] &=
-\int {\rm d}^d x \,\sqrt{g} \Bigl\{ \alpha_1 R_{\mu\nu\rho\sigma}^2  + \alpha_2 R^2
+ I_{2,\mu\nu} E[g]^{\mu\nu}
\Bigr\}
\,,
\\
I_{2,\mu\nu} &= \alpha_3 R g_{\mu\nu} + \alpha_4 R_{\mu\nu} + \alpha_5 \Lambda g_{\mu\nu}\,.
 \end{split}
\end{equation}
There is some freedom in the way in which the tensor $I_{2,\mu\nu}$ is parametrized,
as itself can contain the equations of motion, however
the important points are that all of the terms of \eqref{eq:hd-operators-no-eoms} appear, and that $I_{2,\mu\nu}$ is dimension two and decouples on-shell.
For convenience, we introduce the shorthand notation
\begin{equation}\label{eq:hd-operators-shorthand}
\begin{split}
S_{\rm hd}[g] &= \alpha \cdot {\cal R}_2 [g]\,,
\end{split}
\end{equation}
where $\alpha$ and ${\cal R}_2 [g]$ are interpreted as vectors in the five-dimensional spaces of couplings and dimension four operators, respectively.
The explicit basis can be written down easily by inserting $I_{2,\mu\nu}$ and $E[g]^{\mu\nu}=G^{\mu\nu}+\Lambda/2 g^{\mu\nu}$ in $S_{\rm hd}[g]$.

In this section and from now on
we restrict our attention to the gauge choice $\beta=1$ (i.e., $\delta\beta=0$) in \eqref{eq:dedonder}, which will leave us with only $\lambda$ in \eqref{eq:linear-split} to test parametric dependencies. An important point: notice that our ability
to write down the complete basis \eqref{eq:hd-operators} hinges on the fact that we are working with an arbitrary dimensionality $d$, while for specific dimensions the basis might collapse in two or less terms, suggesting the fact that some operators are evanescent (e.g., in $d=2$ we have that both $R_{\mu\nu}$ and $R_{\mu\nu\rho\theta}$ can be expressed in terms of $R$ and the metric, while in $d=4$ one combination is the topological Euler term). We return on this point when presenting the spectrum of scaling operators for the special cases $d=2,3$ in Sect.~\ref{sect:special-cases}.

\subsection{Path-integral of composite operators}

The bare couplings $\alpha_i$ act as sources for the (integrated) higher derivative operators.\footnote{%
It would be possible to source the local composite operators by replacing $\alpha_i\to \alpha_i(x)$, which would require an analysis more akin to the local renormalization group \cite{Shore:1986hk}. This would be more difficult, but it could have implications for future applications of asymptotic safety, especially for studying observables. We hope that this is addressed in future research.}
If we include $S_{\rm hd}$ in \eqref{eq:pi}, we obtain a functional,
\begin{equation}\label{eq:pi2}
\begin{split}
 Z[j;\alpha] &= \int {\rm D}h \, {\rm e}^{-S_E[\overline{g}+h]-S_{\rm hd}[\overline{g}+h]
 -S_{\rm gf}-S_{\rm gh} + j\cdot h}\,,
\end{split}
\end{equation}
which depends on the new sources and derivatives wrt $\alpha$ realize the expectation values of the operators in \eqref{eq:hd-operators}
\begin{equation}\label{eq:expectation-composite}
\begin{split}
 \langle {\cal R}_2\rangle = -\frac{\partial}{\partial \alpha} Z[j;\alpha]_{j=\alpha_i=0} \,.
\end{split}
\end{equation}
Since the perturbative expansion is based on Newton's constant, the additional higher derivative operators can be treated as composites and it is sufficient to
determine their \emph{linear} renormalization to determine \eqref{eq:expectation-composite}. {
In other words, the renormalization of $Z[j;\alpha]$ is ultimately necessary
only in the limit \eqref{eq:expectation-composite}, which gives the expectation values of the composite operators. Further orders would be useful to construct expectation values of two-point functions of the composite operators summarized in Eq.~\eqref{eq:hd-operators-shorthand}.
Notice also that it is important to regard the composite operators
as ``perturbations'' of the renormalized Einstein-Hilbert action and, by extension, of the asymptotically safe fixed point because,
if they were allowed to dominate the dynamics, then the theory would be governed by
a four-derivative propagator and essentially become higher-derivative gravity, which is a different universality class perturbatively renormalizable in $d=4$, rather than $d=2$.

The new counterterms needed to renormalize \eqref{eq:expectation-composite} can be easily
deduced by including the Hessian $\alpha \cdot \delta^2{\cal R}_2$ of the new contributions \eqref{eq:hd-operators}
in Eq.~\eqref{eq:gamma} through the replacement ${\cal O}_h^\lambda \to {\cal O}_h^\lambda+\alpha \cdot \delta^2{\cal R}_2$ (recall that we have now set $\beta=1$) and by expanding. Since we only need the linear order in the sources $\alpha_i$
for \eqref{eq:expectation-composite}, we deduce the first contribution
\begin{equation}\label{eq:composite-part1}
\begin{split}
- \alpha \cdot \frac{1}{2}\Tr\Bigl\{\frac{\delta^2{\cal R}_2}{\delta h_{\mu\nu}\delta h_{\rho\sigma}}{\cal G}_{h,\mu\nu\rho\sigma}\Bigr\}_{\langle h\rangle=0}\,,
\end{split}
\end{equation}
where ${\cal G}_h$ is the Green function of ${\cal O}^\lambda_h$.
By construction the renormalization of composite operators \eqref{eq:expectation-composite} is always linear
in the couplings of the composite operators themselves in dimensionless rg schemes;
further orders must be computed only if higher-point functions are needed.

The Green function admits an integral representation in terms of the heat-kernel proper time
\begin{equation}\label{eq:propertime-rep-green}
{\cal G}_{h,\mu\nu\rho\sigma} = \int_0^\infty{\rm d}s\;{\rm e}^{-s{\cal O}^\lambda_h}\,,
\end{equation}
and we suppressed the indices on the rhs for brevity.
Inserting \eqref{eq:propertime-rep-green} in the functional trace \eqref{eq:composite-part1}, we find 
\begin{equation}
 \Tr\Bigl\{\frac{\delta^2{\cal R}_2}{\delta h_{\mu\nu}\delta h_{\rho\sigma}}{\cal G}_{h,\mu\nu\rho\sigma}\Bigr\}=
 \int{\rm d}s\;\Tr\left\{\frac{\delta^2{\cal R}_2}{\delta h_{\mu\nu}\delta h_{\rho\sigma}}{\rm e}^{-s{\cal O}^\lambda_h}\right\}\,,
\end{equation}
which must be computed in order to find the new counterterms.

}

It is less clear, however, what is the correct argument of the lhs of Eq.~\eqref{eq:expectation-composite}.
The important point now is to recall the discussion that led to \eqref{eq:gamma-renormalized} and motivates the use of the metric to $\tilde{g}_{\mu\nu}$ in \eqref{eq:ren-metric}, which replaces $\overline{g}_{\mu\nu}$ as the natural argument of the renormalized effective action. Combining the redefinition with \eqref{eq:composite-part1} and, using the notation $J^\lambda_{\mu\nu}\equiv J^{1, \lambda}_{\mu\nu}$, we have
\begin{equation}\label{eq:composite-complete}
\begin{split}
\Delta\Gamma=
S_{\rm hd}
- \alpha \cdot \Tr\Bigl\{\frac{1}{2}\frac{\delta^2{\cal R}_2}{\delta h_{\mu\nu}\delta h_{\rho\sigma}}{\cal G}_{h,\mu\nu\rho\sigma}
+\frac{GJ^{\lambda}_{\mu\nu}}{\epsilon}\frac{\delta{\cal R}_2}{\delta h_{\mu\nu}}
\Bigr\}\,,
\end{split}
\end{equation}
to be evaluated at $\langle h\rangle=0$, which is a functional of $\tilde{g}_{\mu\nu}$. It is the rhs of \eqref{eq:composite-complete} that must be made finite by subtraction and renormalization of $\alpha_i$ to obtain the contribution $\Delta \Gamma_{\rm ren}$ to \eqref{eq:gamma-renormalized}. Similar previous applications of the composite operators, e.g., Refs.~\cite{Becker:2019fhi} and \cite{Houthoff:2020zqy}, overlooked
the second term in parentheses because it was not clear there what the renormalized metric {in the limit $\langle h_{\mu\nu}\rangle=0$} was, but we find that the metric is essential to eliminate the parameter $\lambda$. The essential point is that if the second term of 
\eqref{eq:composite-complete} is not included then the results depend on the scheme, even on-shell, and thus their physical meaning is unclear.

\subsection{Regularization of composite insertions}

The difficult part of the computation of \eqref{eq:composite-complete}
is the trace involving the Green function. The second functional derivative $\delta^2{\cal R}_2$ is a differential operator with up to four covariant derivatives of the background metric that acts on the Green function ${\cal G}_h$.
One can follow the same strategy as the covariant insertion that appeared in \eqref{eq:gamma-h-contribution} and express
${\cal G}_h = \int_0^{\infty} \exp(-s {\cal O}^{\lambda}_h)$. Thus the contribution reduces again to an insertion of a differential operator,
$\delta^2{\cal R}_2$, acting on the heat kernel.
The result admits an asymptotic expansion that schematically is
\begin{equation}
\begin{split}
\alpha \cdot \Tr\, \delta^2{\cal R}_2 \, {\rm e}^{- s {\cal O}^{\lambda}_h} \sim \frac{1}{s^{d/2+2}} \Bigl\{
&
\tilde{a}_0(\overline{g};\lambda,d) + s \, \tilde{a}_1(\overline{g};\lambda,d)
\\&
+ s^2 \, \tilde{a}_2(\overline{g};\lambda,d) +\cdots
\Bigr\} \,,
\end{split}
\end{equation}
not unlike \eqref{eq:sdw-expansions-schematically}, the coefficients of which are known \cite{Groh:2011dw,Jack:1983sk}.
{The discussion on how to regularize the functional traces of the previous section applies without change, so we continue $d$ in the leading power leaving the coefficients untouched
\begin{equation}\label{eq:sdw-expansions-schematically-regulated-hd}
\begin{split}
\alpha \cdot \Tr\, \delta^2{\cal R}_2 \, {\rm e}^{- s {\cal O}^{\lambda}_h}|_{\rm reg} \sim \frac{1}{s^{3+\epsilon/2}} \Bigl\{
&
\tilde{a}_0(\overline{g};\lambda,d) + s \, \tilde{a}_1(\overline{g};\lambda,d)
\\&
+ s^2 \, \tilde{a}_2(\overline{g};\lambda,d) +\cdots
\Bigr\} \,,
\end{split}
\end{equation}
and extract the $1/\epsilon$ poles by inserting \eqref{eq:sdw-expansions-schematically-regulated-hd} into \eqref{eq:composite-complete} to determine the Green function.
The salient features of the heat kernel coefficients are given in Appendix~\ref{sect:appendixHK}.
}

As expected, the computation of \eqref{eq:composite-complete} reveals a result that is expressed in a ``redundant'' basis, because, besides the dimension-two background curvatures, we have also introduced the dimensionful parameter $g_0$ from the original bare action \eqref{eq:einstein-hilbert}. The rhs of \eqref{eq:composite-complete} is thus expressed in a basis that includes
contractions of the equations of motion $E[\overline{g}]^{\mu\nu}$, which should be eliminated by going on-shell, $E[\overline{g}]^{\mu\nu}=\overline{G}_{\mu\nu}+\frac{g_0}{2g_1}\overline{g}_{\mu\nu}=0$
(this is our ``quantum'' equation of motion at the leading order, which would receive corrections beyond one loop).

To make explicit the role of the equations of motion, we write the counterterm action, always in the limit $\langle h_{\mu\nu}\rangle=0$ in analogy with Eqs.~\eqref{eq:sct-comp} and \eqref{eq:hd-operators}, as
\begin{equation}
\begin{split}
 \Delta\Gamma_{\rm div} &= -\frac{\mu^{\epsilon}}{\epsilon}\int
 \Bigl\{B_{d,1} \overline{R}_{\mu\nu\rho\sigma}^2+B_{d,2} \overline{R}^2
 +J^\lambda_{2,\mu\nu} E[\overline{g}]^{\mu\nu}\Bigr\}\,,
 \end{split}
\end{equation}
where $B_{d,1}$ and $B_{d,2}$ are $d$-dependent linear combinations of $\alpha_i/g_1$, while $J^\lambda_{2,\mu\nu}$ is a new source, also linear in $\alpha_i/g_1$, which differs from $J^{\beta,\lambda}_{\mu\nu}$ of the previous section because it is dimension two, for example.
The most important points, that we checked explicitly in our computation, are that the renormalization $J^\lambda_{2,\mu\nu}$ depends on the parameter $\lambda$ because it is an off-shell quantity,
instead the constants 
$B_{d,1}$ and $B_{d,2}$ are independent of the parametrization, implying that we can safely go on-shell even when renormalizing
the composite operators.

Given our choice of how to go on-shell, it is natural at this stage to subtract the divergences $B_{d,i}$ by renormalizing the couplings $\alpha_i$ for $i=1,2$ in \eqref{eq:hd-operators}, and to subtract the divergence $J^\lambda_{2,\mu\nu}$ by renormalizing $I_{2,\mu\nu}$,
i.e., the couplings $\alpha_i$ for $i=3,4,5$, also in \eqref{eq:hd-operators}.
The renormalization is linear by construction, because we are treating the higher derivative operators as composites.
Upon subtraction the renormalized couplings, which by abuse of notation we still denote with $\alpha=\{\alpha_i\}$, obey renormalization group equations linear in themselves, $\beta_i = \mu\partial_\mu \alpha_i = \gamma_{ij}\alpha_j=(\gamma\cdot \alpha)_i$,
but the operators are subject to mixing through a $5\times 5$ matrix.
{All the building blocks to determine $\gamma_{ij}$ are given in Appendix~\ref{sect:appendixHK}.
}

The Callan-Symanzik equation for the renormalized expectation values $\langle {\cal R}_2\rangle_{\rm ren}$ is
\begin{equation}
\begin{split}
\label{eq:CallanSymanzik}
\left(\mu\frac{\partial}{\partial \mu} + \beta_{G}\frac{\partial}{\partial G}+\gamma \,\cdot \right)\langle {\cal R}_2\rangle_{\rm ren} = 0\,.
\end{split}
\end{equation}
At the ultraviolet fixed point the critical properties of the spectrum
come from the diagonalization of the matrix $\gamma$.
We use the standard convention of expressing the scaling exponents
from the analysis of the dimensionless couplings (in units of the renormalization group scale),
${\tilde{\alpha}}_i = \mu^{2-\epsilon} \alpha_i$, with beta functions ${\tilde{\beta}}_i={\tilde{\gamma}}_{ij}{\tilde{\alpha}}_j$, because they are operators of dimension $4-d=2-\epsilon$.

The constants $B_{d,1}$ and $B_{d,2}$ are naturally interpreted as the renormalizations of $\alpha_1$ and $\alpha_2$, as we have done with Newton's constant in the previous section, and they do not depend on the parametrization $\lambda$. The renormalized couplings $\{{\alpha}_3, {\alpha}_4, {\alpha}_5\}$, instead, have trivial beta functions, ${\tilde{\beta}}_i=(4-d){\tilde{\alpha}}_i$, corresponding to the fact that we eliminated three operators in the counterterm Lagrangian using the equations of motion.
The corresponding scaling exponents are thus precisely what we expect from traces and contractions of the equations of motion, so we disregard them.
The interesting beta functions are given by
\begin{align}
\begin{split}
&\tilde{\beta}_1 = (2-\epsilon)\tilde{\alpha}_1+B_{d,1}\,, \quad
\tilde{\beta}_2 = (2-\epsilon)\tilde{\alpha}_2+B_{d,2}\,.
\end{split}
\end{align}
Even though the complete $5\times 5$ mixing matrix $\tilde{\gamma}$ is not block diagonal (there is mixing with the equations of motion), we have directly checked that its remaining two eigenvalues are not affected by the mixing of $\tilde{\alpha}_1$ and $\tilde{\alpha}_2$ with the redundant sector: we can thus extract the scaling exponents from the $2 \times 2$ minor matrix $\hat{\gamma}$
\begin{equation}
\begin{split}
\left(\begin{matrix}
{\tilde{\beta}}_1 \\
{\tilde{\beta}}_2
\end{matrix}\right)
&=
\left(\begin{matrix}
{\hat{\gamma}}_{11} & {\hat{\gamma}_{12}} \\
{\hat{\gamma}_{21}} & {\hat{\gamma}}_{22}
\end{matrix}\right)\left(\begin{matrix}
{\tilde{\alpha}}_1 \\
{\tilde{\alpha}}_2
\end{matrix}\right)\,,
\end{split}
\end{equation}
where $B_{d,i}=\hat{\gamma}_{i,j} \tilde{\alpha}_j$ are restricted to $i,j=1,2$.
The minor is equivalent to not only removing rows and columns of the redundant couplings, but also setting the redundant couplings to zero from the very beginning, as would be required by a truly essential scheme.

Substituting the value $\epsilon=d-2$ with the same logic as in the previous section,
the components of $\hat{\gamma}$ are
\begin{equation}\label{eq:gammaMatrix}
\begin{split}
\hat{\gamma}_{11} &= (4-d)+\frac{1}{4\pi}\frac{(4-d) p_1(d)}{360(d-2) } G\,,\\
\hat{\gamma}_{12} &=  -\frac{1}{4\pi}\frac{(d-1)(4-d)}{180 } G\,,\\
\hat{\gamma}_{21} &= -\frac{1}{4\pi} \frac{p_2(d)}{720 d(d-2)} G\,,\\
\hat{\gamma}_{22} &=  (4-d) + \frac{1}{4\pi}\frac{p_3(d)}{360 d }G \,,
\end{split}
\end{equation}
with the polynomials
\begin{equation}
\begin{split}
p_1(d) =& \, d^4-35d^3+636d^2-718d-124\,,\\
p_2(d) =& \, 5d^6-47d^5+28d^4+198d^3-2416d^2 \\
& \, +14024d-13472\,,\\
p_3(d) =& \, 30d^4-85d^3-567d^2+1710d-1928\,.
\end{split}
\end{equation}
The above result has been confirmed independently using the essential renormalization
group approach discussed in Ref.~\cite{Baldazzi:2021orb}.\footnote{%
We are very grateful to K.~Falls and R.~Ferrero for private communication on this point.
}
The components of $\hat{\gamma}$ have to be evaluated at the fixed point $G=G^*$ given in \eqref{eq:uv-fp} to deduce the scaling properties of the operators in the uv. One may wonder about the singularity appearing in Eq.~\eqref{eq:gammaMatrix} for $d=2$, which is a reminder of the kinematical pole from \eqref{eq:Kinv}. Note that, at the uv fixed point, the value of the Newton's constant cancels the poles. Moreover, as we will see shortly, in $d=2$ there is a degeneracy of the higher derivative operators and a careful analysis shows that the flow is actually well-defined.

\begin{figure}
 \includegraphics[width=7cm]{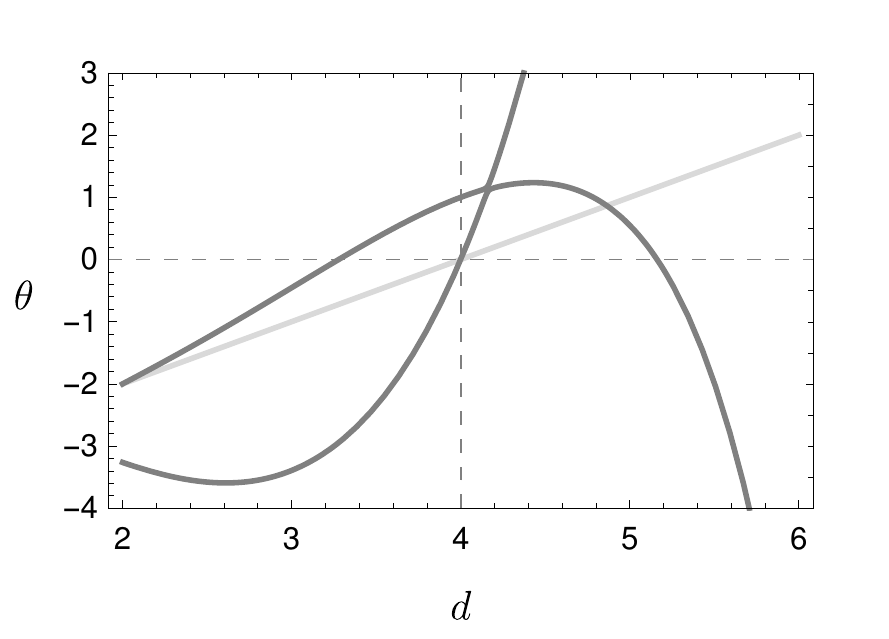}
 \caption{Critical exponents of the essential higher-derivative composite operators as functions of $d$.
 The critical exponent that starts as $-2$ in $d=2$ (the canonical value)
 and becomes positive before $d=4$ reduces to the ``essential'' $R^2$ operator in the physical case $d=4$. The other critical exponent roughly corresponds to the square of the Riemann tensor: it has a noncanonical value in $d=2$ because of the fact that the continuation of the Riemann tensor to two dimensions is singular, and it becomes zero in $d=4$ by interpolating with the topological four-dimensional Euler term. The continuation is helpful to understand the onset of a higher dimensionality in the uv critical surface, but for the special limits $d=2,3$ extra constraints in the curvatures must be taken into account, as explained in Sect.~\ref{sect:special-cases}. The light colored straight line corresponds to canonical scaling.}
 \label{fig:critical-exps}
\end{figure}

The explicit expression for the eigenvalues and of the eigenvectors of $\hat{\gamma}$ for a general $d$ is unwieldy and not particularly illuminating, so we plot critical exponents as functions of $d$ in Fig.~\ref{fig:critical-exps}.
It is interesting to notice that, in the physical case $d=4$, one of the exponents vanishes exactly because of the appearance of the Euler density which, on-shell, coincides with the squared Riemann tensor. The eigensystem of $\hat{\gamma}$ in $d=4$ reduces to the following combination of eigenvalues $\{-\theta_1, -\theta_2\}$ and eigendirections $\{v_1, v_2\}$
\begin{align}
&\theta_1=0\,, && v_1 = (2,1) \,; \\
&\theta_2=1\,, &&  v_2 = (0,1)\,.\nonumber
\end{align}
We deduce some important facts from the analysis of the critical properties. According to the extrapolation of our perturbative analysis,
asymptotic safety in $d=4$ dimensions is expected to have only one additional uv relevant direction taking the form of the $R^2$ operator, and an exactly marginal one which is the topological term.
Most importantly, other operators do not contribute nontrivially to the spectrum, but rather scale trivially as one would expect from combinations of the equations of motion.
In other words, asymptotic safety begins to be truly predictive (i.e., there are uv irrelevant directions) beyond the quadratic order in the curvatures.

\section{The special cases $\bm{d=2}$ and $\bm{3}$} \label{sect:special-cases}

It is important to discuss the other integer values $d=2$ and $3$, because they are special limits that need to be treated with care. In fact, in these cases we have less than three independent contractions of the Riemann tensor with itself, so the basis given in \eqref{eq:hd-operators} is redundant and we need some further manipulations to extract the actual critical exponents. In $d=2$ one always has
\begin{equation}
\begin{split}
R_{\mu\nu\rho\sigma} =& \frac{R}{2}\left(g_{\mu\rho}g_{\nu\sigma}-g_{\nu\rho}g_{\mu\sigma}\right)\,, \quad
R_{\mu\nu} = \frac{R}{2}g_{\nu\sigma}\,,
\end{split}
\end{equation}
implying that the squares of both the Riemann and Ricci tensors are proportional to $R^2$. Therefore, if we wish to extract the correct result from our computations, we need to take into account the fact that the essential couplings $\{\tilde{\alpha}_1, \tilde{\alpha}_2\}$ are not independent. Since the equations of motion are themselves singular in $d=2$, the best strategy is to consider the general $d$-dependent result given in the previous section,  and properly project to a linear combination of beta functions.
For the projection, we are interested in the coupling $\tilde{\alpha}_2$, but must also take into account that the square of the Riemann tensor is proportional to $R^2$ in the limit. Consequently we have
\begin{equation}
\tilde{\beta}_2^{d=2} = \left(\tilde{\beta}_1+\tilde{\beta}_2\right)^{d\to 2}_{\tilde{\alpha}_1=0}\,,
\end{equation}
where on the rhs there are the general $d$-dependent beta functions computed before.
The complete mixing matrix $\gamma$ becomes $3\times 3$ and, as one might expect, in this limit the scaling of $\tilde{\alpha}_2$ becomes the classical one: $\theta=2$. The same remains true for $\tilde{\alpha}_4$ and $\tilde{\alpha}_5$ since we use the corresponding operators to go on-shell.
One might be surprised that the scaling of $\tilde{\alpha}_4$ does not vanish, as would be suggested by the fact that its conjugate operator becomes the topological $2$-dimensional Euler term. However, the equations of motions are discontinuous in this limit and imposing them for a generic $d$ hides the identification with the Euler term.

In $d=3$ there is one relation among the curvatures
\begin{equation}
\begin{split}
 R_{\mu\nu\rho\sigma}= &\, g_{\mu\rho} R_{\nu\sigma}+g_{\nu\sigma} R_{\mu\rho}-g_{\nu\rho} R_{\mu\sigma}-g_{\mu\sigma} R_{\nu\rho}
 \\&
 -\frac{R}{2}\left(g_{\mu\rho}g_{\nu\sigma}-g_{\nu\rho}g_{\mu\sigma}\right)\,,
 \end{split}
\end{equation}
which implies
\begin{equation}
R_{\mu\nu\rho\sigma}R^{\mu\nu\rho\sigma} = 4 R_{\mu\nu}R^{\mu\nu}-R^2\,.
\end{equation}
Following the same strategy as the previous case, we can forget about the first coupling, $\tilde{\alpha}_1=0$, and the beta function for $\tilde{\alpha}_2$ becomes
\begin{equation}
\tilde{\beta}_2^{d=3} = \left(-\tilde{\beta}_1+\tilde{\beta}_2\right)^{d\to 3}_{\tilde{\alpha}_1=0}\,,
\end{equation}
which also depends on the general $d$-dependent beta functions on the rhs.
At the uv fixed point,
we find the nontrivial critical exponent $\theta=-\frac{49}{108}$,
while the rest of the spectrum has canonical scaling (in $d=3$ the canonical scaling of higher derivative operators is $-1$).

The analysis of both special cases confirms that, in the range $2<d<4$,
it is the operator $R^2$ that influences the dimensionality of the uv critical surface in the space of couplings. In particular, also in qualitative agreement with the general plot of Fig.~\ref{fig:critical-exps}, in $d=3$ the critical surface is two-dimensional while in $d=4$ it is three-dimensional. In the analytic continuation, the dimensionality of the critical surface changes somewhere above $d=3$.
The fact that the $R^2$ operator is a uv relevant deformation of the critical surface is not new in the literature of asymptotic safety, however,
our analysis clearly shows that the other directions are not part of an essential scheme that goes on-shell
and further (ir)relevant deformations should be looked for in the spectrum of operators with more than four derivatives.

\section{Conclusions}\label{sect:conclusions}

In this paper we have pushed forward the program initiated in Ref.~\cite[Sect.~IV]{Martini:2021slj} of using dimensional regularization close to two dimensions to discuss the asymptotically safe fixed point of gravity \emph{above} $d=2$.
The technical environment is the one of a background field computation, performed covariantly thanks to heat kernel methods, that reveals the existence of a uv completion of standard Einstein-Hilbert gravity. The completion is based on an ultraviolet fixed point for Newton's constant $G$ that exists, formally, as an expansion $G^*\sim (d-2)$. We have given indication on what the $d$-dependence of the fixed point is even outside the regime $d-2\ll 1$, thanks to the fact that, at least at the leading order in perturbation theory, $d$ can be continued in a way similar to the parameter $N$ of $SU(N)$ gauge theories. At face value the results suggest that in $d=4$ the fixed point exists and is scheme- and gauge-independent, as confirmed independently by Ref.~\cite{Falls:2017cze}.

We are aware of the limitations of perturbation theory when applying the results to the physically interesting limit $d=4$, which is why we recommend our approach
as one that integrates and complements the results coming from the use of Wilsonian methods \cite{Reuter:1996cp}.

There are two main results of our analysis. The first is that we can deduce that a renormalization of the {(fluctuations of the)} metric is necessary to interpret the renormalized effective action as a proper functional of the metric that is also independent of scheme and gauge. The renormalization is somewhat nontrivial and plays an important role for consistency of further applications, such as the computation of the rg equations of expectation values of composite operators.
We informally refer to the renormalized metric as ``the true quantum metric'' of quantum Einstein's gravity.
This renormalization is a test bed for future directions, in particular the much needed extension to two loops of our present analysis, because the nonlinear renormalization of the metric is expected to play a crucial role when subtracting higher order dimensional poles (in a way similar to other nonlinear theories, such as nonlinear sigma models \cite{Howe:1986vm,Osborn:1987au}).

The first result fully clarifies the counterterms structure of Ref.~\cite{Martini:2021slj} and consolidates the importance of going on-shell to prove the full scheme- and gauge-independence of the results. As far as we know, this is largely ignored in the Wilsonian-based literature of asymptotically safe gravity, because an off-shell effective action is deemed satisfactory for physical applications, however such a Wilsonian action is manifestly gauge-dependent, making its careless application potentially suspicious. We would like to see this problem addressed in the Wilsonian framework, and we believe that obtaining an equivalent notion of a true quantum metric is key.

The second main result of this paper is the analysis of higher derivative operators, quadratic in the curvatures, on the premise of the first result.
The analysis is motivated by the need to compare our approach with
the Wilsonian ones, where the spectrum of (ir)relevant operators is generally provided, although often in some approximation, to establish the finiteness of the uv critical surface of the fixed point.

Several facts become clear through the analysis, but the most important ones are that the use of the true quantum metric is necessary to fully subtract the divergences and write everything in terms of renormalized quantities. Furthermore, the spectrum of quadratic operators needs to go on-shell too, as with the rest of the effective action. We find that there is only one physically interesting scaling operator, while off-shell approaches would find two or more, but they would be scheme- and gauge-dependent. The scaling operator corresponds to the square of the Ricci scalar, which is not surprising if the geometrical properties of the square tensors are kept in mind.

One interesting aspect of the analysis is that fixed point and critical exponents are computed as $d$-dependent functions, using essentially the same method of continuing the parameters as done in gauge theories. If we trust the continuation from $d=2$ to $d=4$ of the result, which is of course an extrapolation of perturbation theory, we see that the critical exponent of the $R^2$ operator starts irrelevant in $d=2$ (the coupling has negative canonical dimension), but becomes relevant \emph{before} $d=4$.
This fact confirms the finding of the Wilsonian approaches and, we believe,
increases their intrinsic value, as we now expect that the corresponding relevant $R^2$ deformation of the critical effective action is truly physical on-shell.
All of these considerations, in our eyes, give an added value to the past studies of $R+R^2$ truncations of the renormalization group, e.g., \cite{Bonanno:2013dja}.

Our work has two natural directions at this stage. On the one hand, it remains to be proven that our renormalization procedure continues to work beyond one loop (and especially at two loops, where some of its features would be really tested). In order for the procedure to work, we must be able to treat the dimensionality that is continued for dimensional regularization and the $d=g^\mu{}_\mu$ of spacetime independently in a consistent way. The composite operators are a sort of first test of this fact, given the many nontrivial cancellations that occurred in this paper. On the other hand, it would be interesting to expand our approach to cubic curvature operators, because recent work has pushed forward the idea that among all cubic operators there is a uv relevant direction in $d=4$ \cite{Kluth:2020bdv}, so it would be promising to see this feature confirmed by our approach.

\smallskip

\paragraph*{Acknowlegments.}
We are grateful to D.~Anselmi,
K.~Falls, R.~Ferrero, H.~Gies, D.~Litim, C.~Pagani and R.~Percacci for useful discussions.

\appendix

\section{Gravity and dimensional regularization}\label{sect:appendix}

A justified skepticism, which emerged during this paper's revisions, is on the regularization scheme that we adopt and discuss at length in Sect.~\ref{sect:introduction}. It is related to the fact that the Einstein term is topological in $d=2$, among other things. Here we give our take on
this point by summarizing the main conclusions of Ref.~\cite{Martini:2021slj}, framing them in such a way that the above skepticism is addressed.

As said, in $d=2$ the integral of the scalar curvature is topological, i.e., it is the two-dimensional Euler characteristic. This means that, strictly speaking, the action \eqref{eq:einstein-hilbert} has trivial equations of motion. At the same time, in two dimensions the metric is completely determined by its conformal class, at least locally. The main past strategy to construct
a theory of quantum gravity in $d=2$ has thus been to change \eqref{eq:einstein-hilbert} so that a dynamics is given to the conformal mode. This can be achieved, for example, by coupling the theory to a dilaton, as done in string theory, which requires the cancellation of the conformal anomaly for consistency.
This procedure works very well in $d=2$, because it is always possible
to cancel the conformal anomaly, thanks, in fact, to the topological nature of
the Einstein terms. However, as shown explicitly in Ref.~\cite{Martini:2021slj},
the method cannot be easily continued to $d>2$, because the topological charge
associated with the anomaly acquires a nontrivial running, so it cannot be set to the given value required for the cancellation.

For the above reasons, in Sect.~\ref{sect:graviton-loops} we are prompted to find a different way to analytically continue the theory in $d > 2$, while at the same time using perturbation theory. Notice the following: in $d\neq 2$ the action \eqref{eq:einstein-hilbert} does propagate degrees of freedom, with the case $d=2$ being a discontinuity if the number of components of $g_{\mu\nu}$ is analytically continued naively. As a consequence, if we could use the fact that
$g_{\mu\nu}$ has $d(d+1)/2$ off-shell
components in $d\neq 2$, while, at the same time,
dimensionally regularize $d$ close to $d=2$ for the Feynman diagrams to converge, we would solve all our problems.

In our method, this conundrum is solved by continuing instances of the dimensionality in two separate ways. The geometrical quantities, such as the metric and the curvature tensors, are always to be thought as $d$-dimensional, for $d\neq 2$, implying that 
\eqref{eq:einstein-hilbert} has nontrivial equations of motion, as seen in \eqref{eq:sct-comp}.
In contrast, the measure of covariant Feynman diagrams (that is, the leading power heat kernel as in \eqref{eq:sdw-expansions-schematically}, which plays the role of covariant Feynman parameter), is continued
to $d=2+\epsilon$ dimensions, which allows for nontrivial poles
in dimensional regularization. This can be easily seen in passing from Eq.~\eqref{eq:sdw-expansions-schematically} to Eq.~\eqref{eq:sdw-expansions-schematically-regulated}, where only the leading power of the heat kernel is continued to ensure regularity of the results, while all other instances of $d$ are kept parametrically and survive the computation of the divergent parts of the effective action given in $\Gamma_{\rm div}$.

If one is only interested in the limit $d=4$,
it would be meaningful to think at our procedure as regularizing (the degrees of freedom of) four-dimensional gravity by just continuing the Feynman diagrams close to two dimensions, while keeping the four dimensional structure of the metric intact.
A similar operation was done in Ref.~\cite{Jack:1990pz} to treat quadratic divergences of self-interacting scalars in dimensional regularization.
We indulge once more in the analogy with the continuation in $N$ of the $SU(N)$ Yang-Mills beta function. In our case, $d$ is both the dimensionality of spacetime and a parameter in the local Euclidean/Lorentz gauge group.

At the end of the computation, one can play around with the limit $d\to 2$,
as we also do in this paper, to confirm that the result is genuinely different from the method akin to string theory, which predicts a different beta function for $G$ \cite{Aida:1996zn} if compared with Eq.~\eqref{eq:betaG-JJ}.
However, for us the physically important limit remains the case $d=4$,
in which the metric was always four-dimensional and never two-dimensional to begin with.

As discussed at length in Ref.~\cite{Martini:2021slj}, for the procedure to make complete sense, it is necessary to continue Feynman diagrams below $d=2$ ($d=2+\epsilon$ for $\Re(\epsilon)<0$),
where the conformal mode of the metric is stable. A stronger objection to our method would thus be that we are continuing the regularization to $\epsilon>0$, where the conformal mode becomes unstable and, consequently, it is not clear if we are choosing the correct vacuum for the theory.
For this other objection we have no clear cut solution, however an old idea by Hawking in Ref.~\cite{Hawking:1978jn} comes to the rescue: it can be advocated that the conformal mode of gravity should itself be Wick-rotated for $d>2$, together with time, when passing to the Euclidean signature.
The stability of the conformal mode of quantum gravity may ``simply'' be resolved nonperturbatively, as suggested in the literature of asymptotic safety \cite{Bonanno:2013dja}. It certainly remains an open problem that deserves further investigation.

\section{Heat kernel coefficients}\label{sect:appendixHK}

The traces given in the main text can be computed starting from the asymptotic expansions of the heat kernel for the graviton's differential operator
\begin{equation}\label{eq:hk-graviton}
\begin{split}
 \langle x|{\rm e}^{-s{\cal O}^{\lambda}_h}|y\rangle\sim &\frac{\sqrt{\Delta(x, y)}}{(4\pi s)^{d/2}}{\rm e}^{-\sigma(x, y)/2s}\sum_{n=0}^\infty
 A_n^{\mu\nu\alpha\beta}(x, y)s^n\,,
\end{split}
\end{equation}
and the ghosts' differential operator
\begin{equation}\label{eq:hk-ghosts}
\begin{split}
 \langle x|{\rm e}^{-s{\cal O}^{\lambda}_{\rm gh}}|y\rangle\sim &\frac{\sqrt{\Delta(x, y)}}{(4\pi s)^{d/2}}{\rm e}^{-\sigma(x, y)/2s}\sum_{n=0}^\infty
 B_n^{\mu\nu}(x, y)s^n\,.
\end{split}
\end{equation}
By taking covariant (background) derivatives of these expressions and the coincidence limit $x\to y$, it is possible to derive all the asymptotic expansions given in the main text.
The coincidence limit is enforced on divergences because they are always local, see for example Ref.~\cite{Osborn:1987au}.
Before moving on, notice that all the insertion operators appearing in the main text must be defined after the formal field redefinition
\begin{align}
h_{\alpha\beta}\,\longrightarrow\,h_{\mu\nu}\sqrt{K^{-1}}^{\mu\nu}{}_{\alpha\beta}\,,
\end{align}
which is required to make the operator ${\cal O}^{\beta,\lambda}_h$ in \eqref{eq:hessian-h-full} of Laplace-type when integrating over $h_{\mu\nu}$.

\begin{widetext}

The heat-kernel coefficients of the main text can be computed with some work.
We give their expressions in relation to the bitensors appearing in Eqs.~\eqref{eq:hk-graviton} and \eqref{eq:hk-ghosts} as well as their final form.
The graviton's coefficients are
\begin{align}
\begin{split}
 a_1(\overline{g}; \lambda, d)
 &\equiv  \left.\frac{\sqrt{\Delta(x, y)}}{(4\pi)^{d/2}} A_1{}^{\mu\nu}{}_{\mu\nu}(x, y)\right|_{y\to x}
 \\
 &= \left[d^2(3\lambda-5)-d(3\lambda-7)-12\lambda\right]\frac{\overline{R}}{12(4\pi)^{d/2}}
 +\frac{d\, g_0}{4(4\pi)^{d/2}g_1(d-2)}\left[d(d-1)(2-\lambda)-4(1-\lambda)\right]\,,
 \\
 \hat{a}_1(\overline{g}; \lambda, d)
 &\equiv
 \left.\frac{2}{(4\pi)^{d/2}}A_1{}^{\mu}{}_{\mu}{}^{\nu}{}_{\nu}(x, y)\sqrt{\Delta(x, y)}
 -\frac{2}{(4\pi)^{d/2}(d-2)}\left(\overline{\nabla}^2\overline{g}_{\mu\nu}+d\,\overline{\nabla}_\mu\overline{\nabla}_\nu\right)
 \left(A_0{}^\alpha{}_\alpha{}^{\nu\mu}(x, y)\sqrt{\Delta(x, y)}\right)\right|_{y\to x}\\
 &=-\frac{1}{3(4\pi)^{d/2}}(5d+6\lambda-22)\overline{R}+\frac{g_0}{g_1}\frac{2 d}{(4\pi)^{d/2}}\frac{(d+\lambda-2)}{d-2}\,,
\end{split}
\end{align}
and the ghost coefficients are
\begin{align}
\begin{split}
 b_1(\overline{g}; \lambda, d)
 &\equiv
 \left.\frac{\sqrt{\Delta(x, y)}}{(4\pi)^{d/2}}B_1{}^{\mu}{}_{\mu}(x, y)\right|_{y\to x}
 = \frac{d+6}{6(4\pi)^{d/2}}\overline{R}
 \\
 \hat{b}_1(\overline{g}; \lambda, d)
 &\equiv
 \left.\frac{1}{(4\pi)^{d/2}}B_1{}^{\mu}{}_{\mu}(x, y)\sqrt{\Delta(x, y)}
 -\frac{2}{(4\pi)^{d/2}}\overline{\nabla}_\mu\overline{\nabla}_\nu\left(B_0^{\mu\nu}(x, y)\sqrt{\Delta(x, y)}\right)\right|_{y\to x}
=
 \frac{d+10}{6(4\pi)^{d/2}}\overline{R}\,.
\end{split}
\end{align}
Combining the above coefficients correctly, it is relatively easy to find all the counterterms discussed in Sect.~\ref{sect:introduction}. Alternatively, one could decide to proceed as in Ref.~\cite{Gusynin:1990ek} and consider the full nonminimal operators and compute the counterterms directly.

The divergent part of the expectation values of the composite higher derivative operators is given in terms of the coefficient $\tilde{a}_2(\bar{g};\lambda, d)$. This is considerably more complicate to compute, but the computation can be performed with the aid of {\tt xAct} and {\tt xTras} Mathematica packages \cite{xact,Nutma:2013zea}. The final result has the following form
\begin{align}
\begin{split}
 \tilde{a}_2(\bar{g}; \lambda, d)=&\frac{q_1(\lambda, d)}{\epsilon}R_{\mu\nu\rho\sigma}^2+\frac{q_2(\lambda, d)}{\epsilon}R^2+
 \frac{q_3(\lambda, d)}{\epsilon}R_{\mu\nu}^2+ \frac{q_4(\lambda, d)}{\epsilon}\Lambda^2 + \frac{q_5(\lambda, d)}{\epsilon}R\Lambda\,,
\end{split}
\end{align}
where the functions $q_i(\lambda, d)$ read
\begin{align*}
q_1(\lambda, d) =& -\frac{(4\pi)^{-d/2}}{1440g_1(d-2)^2} (4   {c_1}  (d^6+d^5  (-45 \lambda ^2+180 \lambda -191 )+d^4 (90 \lambda ^2-480 \lambda
   +944 )+2 d^3  (315 \lambda ^2-1020 \lambda +983 )\\
   &-4 d^2  (405 \lambda ^2-1470 \lambda +5212 )-40 d  (27 \lambda ^2-120
   \lambda -925 )+16  (135 \lambda ^2-900 \lambda -662 ) )\\
   &-8   {c_2}  (661 d^4+d^3  (45 \lambda ^2+720 \lambda
   -5109 )+2 d^2  (135 \lambda ^2-1830 \lambda +6554 )-12 d  (45 \lambda ^2-360 \lambda +1003 )\\
   &+8  (45 \lambda ^2-180
   \lambda +302 ) )+  {c_3}  (d^6+d^5  (-45 \lambda ^2+180 \lambda -191 )+6 d^4  (15 \lambda ^2-200 \lambda
   +277 )\\
   &+4 d^3  (135 \lambda ^2+570 \lambda -2624 )-24 d^2  (90 \lambda ^2-70 \lambda -1509 )-16 d (690 \lambda +2843)+32
    (45 \lambda ^2+540 \lambda +238 ) ) )\,,\\
q_2(\lambda, d) =& \frac{(4\pi)^{-d/2}}{576 g_1  (d-2)^2}  (4   {c_1}  (d^6 (5-3 \lambda )^2+d^5  (-63 \lambda ^2+198 \lambda -155 )+2 d^4  (45 \lambda ^2-126 \lambda
   +82 )+d^3  (234 \lambda ^2-924 \lambda +982 )\\
   &-16 d^2  (36 \lambda ^2-153 \lambda +157 )-8 d  (27 \lambda ^2+24 \lambda
   +103 )+32  (18 \lambda ^2-72 \lambda +173 ) )-8   {c_2}  (6 d^5 (3 \lambda -5)\\
   &+d^4  (-9 \lambda ^2-114 \lambda
   +271 )+3 d^3  (21 \lambda ^2+76 \lambda -299 )+d^2  (-36 \lambda ^2+132 \lambda +628 )-36 d  (3 \lambda ^2+36 \lambda
   -55 )\\
   &+16  (9 \lambda ^2+72 \lambda -161 ) )+  {c_3}  (d^6 (5-3 \lambda )^2+d^5  (-63 \lambda ^2+198 \lambda
   -155 )+6 d^4  (18 \lambda ^2-88 \lambda +79 )\\
   &+4 d^3  (27 \lambda ^2+240 \lambda -362 )-24 d^2  (21 \lambda ^2+50
   \lambda -139 )-80 d (6 \lambda +13)+32  (9 \lambda ^2+108 \lambda -134 ) ) )\,,\\
q_3(\lambda, d) =&  -\frac{(4\pi)^{-d/2}}{1440  g_1  (d-2)^2}  (4   {c_1}  (d^6+d^5  (-45 \lambda ^2+180 \lambda -191 )+d^4  (90 \lambda ^2-480 \lambda +944 )+2 d^3
    (315 \lambda ^2-1020 \lambda +983 )\\
    &-4 d^2  (405 \lambda ^2-1470 \lambda +5212 )-40 d  (27 \lambda ^2-120 \lambda
   -925 )+16  (135 \lambda ^2-900 \lambda -662 ) )\\
   &-8   {c_2}  (661 d^4+d^3  (45 \lambda ^2+720 \lambda -5109 )+2
   d^2  (135 \lambda ^2-1830 \lambda +6554 )-12 d  (45 \lambda ^2-360 \lambda +1003 )\\
   &+8  (45 \lambda ^2-180 \lambda
   +302 ) )+  {c_3}  (d^6+d^5  (-45 \lambda ^2+180 \lambda -191 )+6 d^4  (15 \lambda ^2-200 \lambda +277 )\\
   &+4 d^3
    (135 \lambda ^2+570 \lambda -2624 )-24 d^2  (90 \lambda ^2-70 \lambda -1509 )-16 d (690 \lambda +2843)+32  (45 \lambda
   ^2+540 \lambda +238 ) ) )\,,\\
q_4(\lambda, d) =& \frac{d (4\pi)^{d/2}g_1}{64 (d-2)^3}  (\frac{d-1}{g_1}  (\frac{d+2}{g_1}  (4  {c_1}  (d^4 (\lambda -2)^2-6 d^3 (\lambda -2)^2+10 d^2 (\lambda -2)^2+4 d (\lambda -4) \lambda\\
   &-8  (3 \lambda ^2-8 \lambda +4 ) )+8  {c_2}  (d^2  (\lambda ^2-4 \lambda -4 )+d  (-6 \lambda ^2+8 \lambda
   +16 )+4  (\lambda ^2-4 ) )\\
   &+ {c_3}  (d^4 (\lambda -2)^2-6 d^3 (\lambda -2)^2+4 d^2  (3 \lambda ^2-12 \lambda
   +8 )-8 d  (\lambda ^2-4 )-16 (\lambda -2)^2 ) )\\
   &+8  {c_4} (d-2)^2  (d^2 (\lambda -2)-2 d (\lambda -2)-4 \lambda
    ) )+16  {c_5} (d-2)^2  (d^2 (\lambda -2)-d (\lambda -2)-4 \lambda +4 ) )\,,\\
q_5(\lambda, d) =& -\frac{(4\pi)^{d/2}}{96 (d-2)^2}  (\frac{1}{g_1}  (4  {c_1}  (d^6  (3 \lambda ^2-11 \lambda +10 )+d^5  (-15 \lambda ^2+53 \lambda -46 )+2 d^4  (3
   \lambda ^2-10 \lambda +8 )\\
   &+d^3  (78 \lambda ^2-294 \lambda +236 )-4 d^2  (30 \lambda ^2-115 \lambda +104 )+d  (-96
   \lambda ^2+256 \lambda +80 )+48  (3 \lambda ^2-14 \lambda +4 ) )\\
   &-8  {c_2}  (3 d^5 (\lambda -2)+d^4  (-3 \lambda
   ^2-4 \lambda +40 )+d^3  (15 \lambda ^2-9 \lambda -98 )+2 d^2  (6 \lambda ^2-13 \lambda +40 )\\
   &+d  (-48 \lambda ^2+96
   \lambda +56 )+24  (\lambda ^2-2 \lambda -4 ) )+ {c_3}  (d^6  (3 \lambda ^2-11 \lambda +10 )+d^5  (-15
   \lambda ^2+53 \lambda -46 )\\
   &+2 d^4  (6 \lambda ^2-39 \lambda +34 )+12 d^3  (4 \lambda ^2-3 \lambda -8 )-16 d^2  (9
   \lambda ^2-17 \lambda -21 )-16 d (17 \lambda +38)+96  (\lambda ^2+4 ) ) )\\
   &+4  {c_4} (d-2)^3  (d^2 (3 \lambda -5)+d
   (17-9 \lambda )-6 (3 \lambda +2) ) )\,.
\end{align*}
In order to obtain the renormalization of the higher derivative composite operators, the above contributions must be combined with the second term of Eq.~\eqref{eq:composite-complete} and expressed in the basis given in Eq.~\eqref{eq:hd-operators}, in which case the scheme's $\lambda$-dependence decouples through the equations of motion. The decoupling of $\lambda$ is obviously rather nontrivial.

\end{widetext}


\end{document}